\documentclass[twocolumn,a4paper,nofootinbib,superscriptaddress,showpacs,eqsecnum,prd]{revtex4-1}

\usepackage{amsmath, amsthm, amsfonts, amssymb, latexsym}
\usepackage{graphicx}
\usepackage{epsfig}

\usepackage{url}

\usepackage[unicode]{hyperref}
\hypersetup{
    unicode=true,          
    pdftitle={Higher-dimensional puncture initial data},
    pdfsubject={Numerical Relativity},
    pdfcreator={LaTeX},          
    pdfproducer={LaTeX},         
    pdfkeywords={General Relativity, Numerical Relativity, Black Holes},
    colorlinks=false,            
  }

\begin{document}

\title{Higher-dimensional puncture initial data}

\author{Miguel~Zilh\~ao}\email{mzilhao@fc.up.pt}
\affiliation{
  Centro de F\'\i sica do Porto,
  Departamento de F\'\i sica e Astronomia,
  Faculdade de Ci\^encias da Universidade do Porto,
  Rua do Campo Alegre, 4169-007 Porto, Portugal
}

\author{Marcus~Ansorg}
\affiliation{
  Theoretisch-Physikalisches Institut, 
  Friedrich-Schiller-Universit\"at Jena, 
  Max-Wien-Platz 1, D-07743 Jena, Germany
}

\author{Vitor~Cardoso}
\affiliation{
  Centro Multidisciplinar de Astrof\'\i sica --- CENTRA,
  Departamento de F\'\i sica, Instituto Superior T\'ecnico --- IST,
  Av. Rovisco Pais 1, 1049-001 Lisboa, Portugal 
}
\affiliation{
  Department of Physics and Astronomy, The University of Mississippi,
  University, MS 38677-1848, USA
}

\author{Leonardo~Gualtieri}
\affiliation{
  Dipartimento di Fisica, Universit\`a di Roma
  ``Sapienza'' \& Sezione,
  INFN Roma1, P.A. Moro 5, 00185, Roma, Italy
}

\author{Carlos~Herdeiro}
\affiliation{
  Departamento de F\'\i sica da Universidade de Aveiro \& I3N, 
  Campus de Santiago, 3810-183 Aveiro, Portugal
}

\author{Ulrich~Sperhake}
\affiliation{
  Institut de Ci\`encies de l'Espai (CSIC-IEEC), Facultat de Ci\`encies, 
  Campus UAB, E-08193 Bellaterra, Spain
}
\affiliation{
  California Institute of Technology,
  Pasadena, CA 91125, USA
}
\affiliation{
  Centro Multidisciplinar de Astrof\'\i sica --- CENTRA,
  Departamento de F\'\i sica, Instituto Superior T\'ecnico --- IST,
  Av. Rovisco Pais 1, 1049-001 Lisboa, Portugal  
}

\author{Helvi~Witek}
\affiliation{
  Centro Multidisciplinar de Astrof\'\i sica --- CENTRA,
  Departamento de F\'\i sica, Instituto Superior T\'ecnico --- IST,
  Av. Rovisco Pais 1, 1049-001 Lisboa, Portugal  
}


\begin{abstract}
We calculate puncture initial data, corresponding to single and binary black holes with
linear momenta, which solve the constraint equations of $D$ dimensional vacuum gravity. The data are generated by a modification of the pseudo-spectral code presented in~\cite{Ansorg:2004ds} and made available as the {\sc TwoPunctures} thorn inside the {\sc Cactus} computational toolkit.
As examples, we exhibit convergence plots, the violation of the Hamiltonian constraint as well as the initial data for $D=4,5,6,7$.
These initial data are the starting point to
perform high-energy collisions of black holes in $D$ dimensions.
\end{abstract}


\maketitle


\section{Introduction}

Numerical relativity in higher-dimensional spacetimes could be a
powerful tool to study a variety of physical concepts, such as the
stability of black hole solutions and their interactions,
as well as for producing phenomenological information of relevance
for TeV gravity scenarios
\cite{Antoniadis:1990ew,Antoniadis:1998ig,Randall:1999ee}. In such
models, the fundamental Planck scale could be as low as 1 TeV.
Thus, high-energy colliders, such as the Large Hadron Collider
(LHC), may directly probe strongly coupled gravitational physics
\cite{Argyres:1998qn,Banks:1999gd,Giddings:2001bu,Dimopoulos:2001hw,
Ahn:2002mj,Chamblin:2004zg}. Indeed, in this scenario, particle collisions could produce black holes
\cite{Giddings:2001bu,Dimopoulos:2001hw}. Moreover, the production
of black holes at trans-Planckian collision energies (compared to the
fundamental Planck scale) should be well-described by using
classical general relativity extended to D dimensions (see
\cite{Kanti:2008eq} and references therein). Numerical simulations
of high-energy black hole collisions in higher-dimensional spacetimes, then,
could give an accurate estimate of the fractions of the collision
energy and angular momentum that are lost in the higher-dimensional
space by emission of gravitational waves; such information would be
extremely important to improve the modeling of microscopic black hole
production, and of the ensuing evaporation phase, which
might be observed during LHC collisions.

The growing interest in dynamical aspects of higher-dimensional
spacetimes led to the development of higher-dimensional numerical
relativity~\cite{Yoshino:2009xp,Sorkin:2009bc,Zilhao:2010sr,Lehner2010}
and to the production of the first black hole collisions in higher-dimensional spacetimes starting from
rest~\cite{Witek:2010xi,Witek:2010az}, and, most recently, of
black holes with initial boost
\cite{Okawa:2011fv}\footnote{These simulations start from
superposed single boosted black hole data without application
of a constraint-solving procedure; our work is motivated in
part by providing initial data for a comparison and thus
calibrating the impact of constraint violations.}.
Aside from their immediate relevance in the context of TeV gravity
scenarios, high-energy collisions of black holes provide fertile
ground for probing strong-field effects of general relativity, such
as cosmic censorship, luminosity limits, hoop type conjectures
and zoom-whirl behaviour
\cite{Sperhake:2008ga, Shibata:2008rq, Sperhake:2009jz, Sperhake2010,Okawa:2011fv}.

The modeling of generic black hole or neutron-star spacetimes in the framework
of Einstein's theory of general relativity requires a numerical treatment
because of the enormous complexity of the equations in the absence of high
degrees of symmetry.  For such numerical modeling, the Einstein equations are
cast as a time evolution or {\em initial value} problem as originally formulated
by Arnowitt-Deser-Misner (ADM) \cite{Arnowitt:1962hi} and reformulated by York
\cite{York1979}.  Numerically generating a solution then consists of two basic
steps: (i) the construction of initial data which satisfy the constraint
equations and represent a realistic snapshot of the physical system under
consideration and (ii) the evolution in time of these initial data. In this work
we will focus on the first step and provide a formalism for constructing initial
data for black hole binaries with nonzero boosts in higher, $D\ge 5$,
dimensional spacetimes.

Most work on the generation of initial data in $3+1$-dimensional general
relativity is based on the York-Lichnerowicz split \cite{Lichnerowicz1944,
  York1971,York1972,York1973} which rearranges the degrees of freedom contained
in the three-metric $\gamma_{ij}$ and extrinsic curvature $K_{ij}$ via a
conformal transformation and a split of the curvature into trace and traceless
part
\begin{equation}
  \label{eq:conf4}
\begin{aligned}
  \gamma_{ij} & = \psi^4 \hat{\gamma}_{ij}\ , \\
  K_{ij} & = A_{ij} + \frac{1}{3}\gamma_{ij} K\ ,
\end{aligned}
\end{equation}
followed by a transverse-traceless decomposition of either the traceless
extrinsic curvature $A_{ij}$ or a conformally rescaled version $\hat{A}_{ij}$
thereof. For details of these {\em physical} or {\em conformal
  transverse-traceless} decompositions as well as the alternative {\em conformal
  thin-sandwich} formalism, we refer the reader to Cook's
review~\cite{Cook:2000vr}, Alcubierre's book~\cite{Alcubierre:2008} and references therein.

One of the main advantages achieved with this decomposition is the decoupling of
the momentum from the Hamiltonian constraint under the additional simplifying
assumptions of conformal flatness, $\hat{\gamma}_{ij}= \delta_{ij}$ and a
constant trace of the extrinsic curvature $K={\rm const}$.  Quite remarkably,
the resulting equations for the momentum constraints admit analytic solutions
describing multiple black holes with nonvanishing spins and linear
momenta~\cite{Bowen:1980yu}. There then remains a single elliptic differential
equation, the Hamiltonian constraint, for the conformal factor $\psi$ which
requires a numerical treatment. By applying a compactification to the internal
asymptotically flat region, Brandt \& Br{\"u}gmann~\cite{Brandt1997} derived a
method of solving the Hamiltonian constraint which is now generally referred to
as the {\em puncture} method.  These puncture data form the starting point for
the majority of numerical simulations using the {\em moving-puncture}
method~\cite{Campanelli2006, Baker2006}. We note that an alternative approach to
evolving the Einstein equations based on the {\em generalized harmonic}
formulation has been implemented with similar
success~\cite{Pretorius:2005gq,Scheel2008}.

As shown by Yoshino {\em et al.}~\cite{Yoshino:2006kc}, the existence of an
analytic solution of the Bowen-York type for the momentum constraints carries over
to the theory of general relativity in higher dimensions.
These authors also solved the Hamiltonian constraint using a finite difference method. 
Here, we will present a generalization of the spectral solver by Ansorg
{\em et al.}~\cite{Ansorg:2004ds}, used inside the CACTUS toolkit~\cite{Allen1999,cactus}, that solves the Hamiltonian constraint
for black hole binaries in $D\ge 5$ dimensions with nonvanishing initial
boost, and preserves the spectral convergence properties observed
in four dimensions. The compactification implemented in this solver
further facilitates direct interpolation of the initial data onto
computational grids of arbitrary size with mesh refinement employed
in state-of-the-art numerical simulations of black hole
spacetimes.  Finally, the initial data are generated in variables
that can be straightforwardly translated to evolution systems
common in numerical relativity,
such as the Baumgarte-Shapiro-Shibata-Nakamura
\cite{Shibata:1995we,Baumgarte:1998te} system in Cartesian form.
In our case, we specifically translate the spectral solution
into the formalism of~\cite{Zilhao:2010sr} which can thus be readily used to
perform high-energy collisions of two black holes in $D$ dimensions.
We complement our study with the analytic perturbative analysis of
boosted single puncture initial data.

This paper is organized as follows. In Sec.~\ref{chap:init_data}, we review
the constraint equations, the Brill-Lindquist~\cite{Brill:1963yv} and
Bowen-York~\cite{Bowen:1980yu} type initial data and introduce an appropriate
coordinate system. In Sec.~\ref{sec3}, we provide the explicit form of the
elliptic equation that must be solved for determining the Bowen-York initial
data that describes a boosted head-on collision in $D$ dimensions. We also
present its dimensional reduction to four spacetime dimensions,
following~\cite{Zilhao:2010sr}. In Sec.~\ref{sec4}, the case of a single black
hole with linear momentum is considered and it is shown that an approximate
analytic solution can be found for {\em all} variables, including the conformal
factor, in the limit of small momentum to mass ratio. In Sec.~\ref{sec5}, we
explain the modifications we have coded to study the case of two black
holes with aligned linear momentum $P/r_S^{D-3}=\pm 0.8$ ($r_S$ is the Schwarzschild radius) and present results for convergence
tests, constraint violation and the solution itself in $D=4,5,6,7$.

\emph{Notation:} In the remainder of this work, early lower case Latin
indices $a,\,b,\,c,\ldots$ extend from $1$ to $D-1$, late lower case
Latin indices $i,\,j,\,k,\ldots$ run from $1$ to $3$ and early upper case
Latin indices $A,\,B,\,C,\ldots$ from $4$ to $D-1$.

\section{Higher-dimensional initial data}
\label{chap:init_data}

The starting point for our discussion is a $(D-1)$-dimensional spacelike
hypersurface $\bar \Sigma$ with induced metric $\bar{\gamma}_{ab}$, and extrinsic curvature $\bar{K}_{ab}$ embedded in a $D$-dimensional spacetime.  By generalizing the ADM decomposition, the spacetime
metric is given in the form
\begin{align}
   ds^2 & = -\alpha^2 dt^2
  + \bar \gamma_{ab} \left(
    dx^a + \beta^a dt
  \right) \left(
    dx^b + \beta^b dt
  \right) \ .
\end{align}
We conformally decompose the spatial metric and extrinsic curvature as
\begin{equation}
  \label{eq:1}
  \begin{aligned}
    \bar  \gamma_{ab} & = \psi^{\frac{4}{D-3}} \hat \gamma_{ab}\ , \\
    \bar K_{ab} & = \psi^{-2} \hat A_{ab} + \frac{1}{D-1} \bar \gamma_{ab} \bar{K}\ ,
  \end{aligned}
\end{equation}
which generalizes the 3+1 dimensional Eq.~(\ref{eq:conf4}). Note that we
here represent the traceless part of the extrinsic curvature by
the conformally rescaled version $\hat A^{ab}$.
We assume that the conformal metric $ \hat
\gamma_{ab}\ $ is flat and impose the maximal slicing condition, $\bar{K}=0$,
which lead to the decoupling of the constraints mentioned above.  With this
choice, the higher-dimensional initial data equations in vacuum
become~\cite{Yoshino:2005ps,Yoshino:2006kc}
\begin{align}
  &  \partial_a \hat A^{ab} = 0 \  , \label{eq:vacuum_momentum} \\
  &  \hat \triangle \psi + \frac{D-3}{4(D-2)} \psi^{- \frac{3D -5}{D-3} }
     \hat A^{ab} \hat A_{ab} = 0\ , \label{eq:vacuum_hamilton}
\end{align}
where $\hat A^{ab} \equiv \hat\gamma^{ac} \hat\gamma^{bd} \hat A_{cd} $, $\hat \triangle \equiv \partial_a \partial^a $ is the flat space
Laplace operator and the first equation holds in a $(D-1)$-dimensional
Cartesian frame $\mathcal{X}^a = (x^1, x^2, \dots, x^{D-1})$.
%

\subsection{Brill-Lindquist initial data}
\label{sec:brill}

For the time-symmetric case $\bar K_{ab} = 0$, Eq.~\eqref{eq:vacuum_momentum} is
automatically satisfied, and Eq.~\eqref{eq:vacuum_hamilton} reduces to the
$D-1$-dimensional flat space Laplace equation,
\begin{equation}
  \label{eq:lappsi}
  \hat \triangle \psi = 0\ .
\end{equation}
For asymptotically flat spacetimes, the conformal factor satisfies the
boundary condition
\begin{equation}
\label{eq:lappsi_conditions}
  \lim_{ r \to \infty} \psi = 1\ .
\end{equation}
and a solution to Eq.~\eqref{eq:lappsi} is given by
\begin{equation}
  \label{eq:psi_BL}
   \psi = \psi_{\mathrm{BL}}=  1 + \sum_{i=1}^{N}
          \frac{\mu_{(i)}}{4 r_{(i)}^{D-3}}\ , \\
\end{equation}
where $r_{(i)} \equiv |r - x_{(i)}|$,
$x_{(i)}$ is the (arbitrary) coordinate location of the $i^{\rm th}$ puncture,
and the mass parameter $\mu_{(i)}$ is related to the horizon radius
$r_{S_{(i)}}$ and the ADM mass $M_{(i)}$ of the $i^{\rm th}$ hole by
\begin{equation}
  \mu_{(i)} \equiv r_{S_{(i)}}^{D-3} \equiv \frac{16\pi
                   M_{(i)}}{\mathcal{A}_{D-2}(D-2)} \ ; \nonumber
\end{equation}
%
here $\mathcal{A}_{D-2}$ is the area
of the unit $(D-2)$-sphere and we have set the $D$ dimensional Newton
constant to unity.

These closed-form analytic data are the $D$-dimensional generalization of
Brill-Lindquist data \cite{Brill:1963yv} and describe a spacetime containing
multiple nonspinning black holes at the moment of time symmetry, \emph{i.e.}~with
vanishing linear momentum.

\subsection{Bowen-York initial data}
\label{sec:bowen}


In order to numerically evolve black holes with nonzero boost, we need to
generalize Brill-Lindquist data to the non time-symmetric case. In four
dimensions, this generalization is given by the Bowen-York extrinsic
curvature, a nontrivial analytic solution of the momentum constraint
Eq.~(\ref{eq:vacuum_momentum}). As shown by
Yoshino {\em et al.} \cite{Yoshino:2006kc}, we can write a solution
of Eq.~(\ref{eq:vacuum_momentum}) describing a spacetime of arbitrary
dimensionality $D$ containing $N$ black holes in the form
\begin{align}
  \label{eq:11}
  \hat A^{ab}_P = \sum_{i = 1}^N \hat A_{P(i)}^{ab}\ ,
\end{align}
where
\begin{widetext}
\begin{align}
  \label{eq:Aab_i}
  \hat A^{ab}_{P(i)} = \frac{4 \pi (D-1)}{(D-2) \mathcal{A}_{D-2} } \frac{1}{r^{D-2}_{(i)} }
  \left(
    n^a_{(i)} P^b_{(i)} + n^b_{(i)} P^a_{(i)} 
    - (n_{(i)})_c P^c_{(i)} \hat \gamma^{ab} + (D-3) n^a_{(i)} n^b_{(i)} P^c_{(i)} (n_{(i)})_c
  \right)\ .
\end{align}
\end{widetext}
%
Here we have introduced $n^a_{(i)} \equiv \frac{x^a - x^a_{(i)} }{r_{(i)}} $
and the parameter $ P^a_{(i)} $ corresponds to the ADM momentum of the
$i^{\rm th}$ black hole in the limit of large separation from all other
holes.


In order to obtain a complete set of initial data we still need to solve
the Hamiltonian constraint \eqref{eq:vacuum_momentum} with $\hat A_{ab}$
given by~\eqref{eq:11}. For this purpose, we follow the standard
decomposition of the conformal factor into a Brill-Lindquist contribution
$\psi_{\rm BL}$ given by (\ref{eq:psi_BL}) plus a regular correction $u$
\begin{align}
  \label{eq:psi-decom}
  \psi = \psi_{\mathrm{BL}} + u\ .
\end{align}
%
Equation~\eqref{eq:vacuum_hamilton} then takes the form
\begin{align}
  \hat \triangle u + \frac{D-3}{4(D-2)} \hat A^{ab} \hat A_{ab}
       \psi^{ -\frac{3D-5}{D-3} } = 0\ . 
  \label{eq:u}
\end{align}
As in $D=4$, the higher-dimensional extension of Bowen-York extrinsic
curvature data can also accommodate angular momentum of the black holes.
In the present work, however, we shall focus on initial data for
nonspinning, boosted black holes only.

\subsection{Coordinate transformation}

In summary, the initial data are determined by (i) the extrinsic curvature
$\bar{K}_{ab}$ obtained by inserting Eq.~\eqref{eq:Aab_i} into~\eqref{eq:11} and
the resulting $\hat{A}_{ab}$ into Eq.~\eqref{eq:1}, and (ii) the spatial $D-1$
metric $\bar{\gamma}_{ab}$ obtained by numerically solving Eq.~(\ref{eq:u}) for
$u$ which gives the conformal factor via Eq.~(\ref{eq:psi-decom}) and the metric
through Eq.~(\ref{eq:1}).

For the numerical solution of Eq.~(\ref{eq:u}), it is convenient to
transform
to a coordinate system adapted to the
generalized axial symmetry $SO(D-2)$ in $D=5$ dimensions and
$SO(D-3)$ in $D\ge6$ dimensions as discussed in Sec.~I C of
Ref.~\cite{Zilhao:2010sr}.
For this purpose we consider the (flat) conformal metric in cylindrical
coordinates
%
  \begin{equation}
    \hat\gamma_{ab}dx^adx^b=dz^2+d\rho^2+\rho^2\left(d\varphi^2+\sin^2\varphi
      d\Omega_{D-4}\right) \ ,
    \label{initialspatial}
  \end{equation}
%
where $d\Omega_{D-4}$ is the metric on the $(D-4)$-sphere.
Observe that $\varphi$ is a polar rather than an azimuthal
coordinate, \emph{i.e.\ }$\varphi\in [0,\pi]$. Next, following \cite{Zilhao:2010sr}, we introduce ``incomplete''
Cartesian coordinates as
\begin{equation}
  x=\rho \cos\varphi \, ,  \qquad y=\rho \sin\varphi \ ,
  \label{inccartesian}
\end{equation}
where $-\infty<x<+\infty$ and $0\le y<+\infty$.
The $D$-dimensional initial data for the spatial metric is then
%
  \begin{equation}
    \bar \gamma_{ab}dx^adx^b=\psi^{\frac{4}{D-3}}\left[dx^2+dy^2+dz^2+y^2
      d\Omega_{D-4}\right]  \ .
    \label{smhigher}
  \end{equation}

We can transform the $D-1$-dimensional Cartesian coordinates
$\mathcal{X}^a = (x^1,
\dots,x^{D-1})$ to the coordinate system $\mathcal{Y}^a = (x,y,z,\xi_1,
\xi_2, \dots, \xi_{D-4})$ with hyperspherical coordinates $\xi_1,\ldots,
\xi_{D-4}$ by
\begin{equation}
  \label{eq:coord-transf}
  \begin{aligned}
    x^1 & = x  & \\
    x^2 & =  y \cos \xi_1 & \\
    x^3 & = z  &\\
    x^4 & =  y \sin \xi_1 \cos \xi_2 \qquad  &(D\ge 6) \\
    x^5 & =  y \sin \xi_1 \sin \xi_2 \cos \xi_3  \qquad  & (D\ge 7) \\
    \vdots \\
    x^{D-3} & = y \sin \xi_1 \cdots \sin \xi_{D-6} \cos \xi_{D-5}  \qquad & (D\ge 7) \\
    x^{D-2} & = y \sin \xi_1 \cdots \sin \xi_{D-5} \cos \xi_{D-4}  \qquad & (D\ge 6) \\
    x^{D-1} & = y \sin \xi_1 \cdots \sin \xi_{D-4}  \qquad & (D\ge 5)
  \end{aligned}\ .
\end{equation}

Without loss of generality, we can always choose coordinates such that the
black holes are initially located on the $z$ axis at $z_1$ and $z_2$
and have momenta of
equal magnitude in opposite directions
$P_{(1)}^a = -P_{(2)}^a$.
Inserting the momenta into Eq.~(\ref{eq:Aab_i}) then provides the conformal
traceless extrinsic cuvature and the differential Eq.~(\ref{eq:u})
which is solved numerically for $u$.

The class of symmetries covered by the formalism developed in
Ref.~\cite{Zilhao:2010sr} includes head-on and grazing
collisions of nonspinning black holes with initial position and momenta
\begin{eqnarray}
  && x_{(1)}^a = (0, 0, z_1,0,\ldots, 0)\,, \quad
     x_{(2)}^a = (0, 0, z_2,0,\ldots, 0) \nonumber \\
  && P_{(1)}^a = (P^x, 0, P^z, 0, \dots, 0) = -P_{(2)}^a \ . \label{eq:pos}
\end{eqnarray}
Note that a nonzero $P^y$ is not compatible with the assumed symmetries.
On the other hand, the $x$-axis can always be oriented such that the
collision takes place in the $xz$-plane. Our formalism therefore
covers general grazing collisions of nonspinning black hole binaries
in $D$ dimensions.

\section{Four dimensional initial data for a general $D$ head-on collision}
\label{sec3}

For illustration and numerical testing, we will in the rest
of this paper discuss in full detail
the case of black holes with momenta in the $z$ direction,
that is, the case given by setting $P^x=0$ in Eq.~(\ref{eq:pos}).
The linear momenta are thus given by
\begin{equation}
  P^a_{(1)} = (0,0,P^z,0,\ldots ,0) = -P^a_{(2)}.
  \label{eq:Pheadon}
\end{equation}
%
The rescaled trace-free part of the extrinsic curvature for such a
configuration is
\begin{equation}
  \label{eq:4}
  \hat A_{ab} = \hat A_{ab}^{(1)} + \hat A_{ab}^{(2)} \ ,
\end{equation}
where $\hat A_{ab}^{(1)}$ and $\hat A_{ab}^{(2)}$ are given by
Eq.~\eqref{eq:Aab_i} with \eqref{eq:pos} and \eqref{eq:Pheadon}.
Using Eq.~\eqref{eq:coord-transf} we can write this in the
coordinate system $\mathcal{Y}^a$ adapted to the spacetime symmetry:
\begin{widetext}
%
\begin{equation}
\label{eq:8}
\hat A_{ab}^{(1)} =
  \displaystyle{
    \frac{4 \pi (D-1) P^z}{(D-2) \mathcal{A}_{D-2}
    (x^2 + y^2 + (z-z_1)^2)^{\frac{D+1}{2}} }
  }
\left(
\begin{array}{c|c}
 \hat a_{i j}^{(1)} & 0  \\
 \hline
 0 & \hat a_{AB}^{(1)}
\end{array}
\right),
\end{equation}
with
%
\begin{equation}
\label{eq:A}
\hat a_{ij}^{(1)} =
\left(
\begin{smallmatrix}
 - \left[-(D-4)x^2+y^2+(z-z_1)^2\right] (z-z_1) & (D-3)  x y
   (z-z_1) &  x \left[x^2+y^2+ (D-2) (z-z_1)^2\right] \\
 (D-3) x y (z-z_1) & - \left[x^2 - (D-4)y^2+(z-z_1)^2\right]
   (z-z_1) &  y \left[x^2+y^2 + (D-2) (z-z_1)^2\right] \\
  x \left[x^2+y^2+ (D-2) (z-z_1)^2\right] &  y \left[x^2+y^2+ (D-2)
   (z-z_1)^2\right] &  \left[x^2+y^2 + (D-2) (z-z_1)^2\right] (z-z_1)
\end{smallmatrix}
\right) \ ,
\end{equation}
\end{widetext}
and
\begin{equation}
\label{eq:9}
  \hat a_{AB}^{(1)} = - y^2 (z-z_1) \left[ x^2 + y^2
                      + (z-z_1)^2 \right] h_{AB}\ ,
\end{equation}
where $h_{AB}$ is the metric on the $(D-4)$-sphere.
The expression for $\hat A^{(2)}_{ab}$ is analogous, but with $z_2$ in place of $z_1$ and
$-P^z$ in place of $P^z$ in Eq.~(\ref{eq:8}).

The formalism developed in \cite{Zilhao:2010sr} for $D$-dimensional spacetimes with $SO(D-2)$
or $SO(D-3)$ isometries describes the spacetime in terms of the
traditional three-dimensional metric $\gamma_{ij}$ and extrinsic
curvature $K_{ij}$ coupled to a scalar field $\lambda$ and its
conjugate momentum $K_{\lambda}$; cf.~Eqs.~(2.14), (2.26) in
\cite{Zilhao:2010sr}. These are the variables evolved in time
and therefore the variables we ultimately wish to construct from
the initial data calculation. For their extraction we first note that
$\gamma_{ij}$, $K_{ij}$ and $K_{\lambda}$ are related to the $(D-1)$-dimensional
metric $\bar{\gamma}_{ab}$ and extrinsic curvature $\bar{K}_{ab}$ by
%
%
\begin{align}
  \bar{\gamma}_{ij} & = \gamma_{ij}\ , \quad
  \bar{\gamma}_{AB} = \lambda h_{AB}\ , \nonumber \\
  \bar{\gamma}_{iA} & = 0\ , \\
  \bar K_{ij} & =  K_{ij}\ , \quad
  \bar K_{AB} = \frac{1}{2} K_\lambda h_{AB}\ , \nonumber \\
  \bar K_{iA} & = 0\ , \quad
  \bar K =  K + \frac{D-4}{2} \frac{K_\lambda}{\lambda}\ .
\end{align}
%
Using these relations and Eq.~(2.14) of \cite{Zilhao:2010sr} we
can express all ``3+1'' variables in terms of those describing the
initial data
\begin{equation}
  \begin{aligned}
    \gamma_{ij} & =  \psi^{\frac{4}{D-3}} 
        \delta_{ij}\ ,\quad 
    %
    \lambda = \psi^{\frac{4}{D-3}} y^2\ , \\
    K_{ij} & = \psi^{-2}(\hat A_{ij}^{(1)} + \hat A_{ij}^{(2)})\ , \quad
    K_\lambda = 2 \psi^{-2} y^2 ( P^{+} + P^{-})\  , \\
    K & = - \frac{(D-4)K_{\lambda}}{2\lambda} \ ,
  \end{aligned}
\end{equation}
where
%
%
%
\begin{equation}
  \begin{split}
P^{+} &\equiv -\frac{4 \pi (D-1) P^z (z-z_1)}{(D-2) \mathcal{A}_{D-2} (x^2 + y^2+ (z-z_1)^2)^{\frac{D-1}{2}}  }\,,\\
P^{-} &\equiv \frac{4 \pi (D-1) P^z (z-z_2)}{(D-2) \mathcal{A}_{D-2} (x^2 + y^2+ (z-z_2)^2)^{\frac{D-1}{2}}  }\,.
\end{split}
\end{equation}
The conformal factor is
\begin{equation}
\begin{aligned}
  \psi = 1 & + \frac{\mu_1}{4 \left[x^2+y^2+(z-z_1)^2 \right]^{(D-3)/2}}
         \\  & +
  \frac{\mu_2} {4 \left[x^2+y^2+(z-z_2)^2 \right]^{(D-3)/2}}  + u \ ,
\end{aligned}
\end{equation}
and $u$ is the solution of the equation
\begin{equation}
\begin{aligned}
   \left( \partial_{\rho\rho} + \partial_{zz} + \frac{D-3}{\rho}
          \partial_{\rho} \right) u
 = \frac{3-D}{4(D-2)} \hat A^{ab} \hat A_{ab} \psi^{ -\frac{3D-5}{D-3} } \ ,
\end{aligned}\label{eq:elleq0}
\end{equation}
where
\begin{equation}
\begin{aligned}
\hat A^{ab} \hat A_{ab}=(\hat A_{ij}^{(1)} + \hat A_{ij}^{(2)}) ( \hat A^{ij}{}^{(1)}+\hat A^{ij}{}^{(2)})\\ 
+ (D-4) ( P^{+} + P^{-})^2\ .
\end{aligned}
\end{equation}
%

Our numerical construction of the function $u$ will be based on the
spectral solver developed in \cite{Ansorg:2004ds}. This solver employs
coordinates specifically adapted to the asymptotic behaviour of $u$
at spatial infinity. In order to investigate this behaviour, we next
consider a single black hole with non-zero linear momentum.

\section{Single puncture with linear momentum}
\label{sec4}

For a single puncture with momentum $P^z$ located at the origin $z=0$,
Eq.~(\ref{eq:Aab_i}) implies
%
\begin{equation}
\begin{aligned}
\label{eq:A2single}
  & \hat A^{ab} \hat A_{ab} = \\ & \frac{16\pi^2 (D-1)^2 }{(D-2)^2
         \mathcal{A}_{D-2}^2 r^{2(D-2)} } P_z^2
         \left[
         2 + D(D-3) \left(\frac{z}{r}\right)^2
\right] \ ,
\end{aligned}
\end{equation}
so that Eq.~\eqref{eq:elleq0} takes the form
\begin{equation}
\begin{aligned}
\label{eq:ueqsingle}
\hat \triangle u 
& +  \frac{8\pi^2 (D-1)^2 (D-3) }{(D-2)^3 \mathcal{A}_{D-2}^2 r^{2(D-2)} }
     P_z^2\times \\
& \times \left[
  1 + \frac{D(D-3)}{2} \left(\frac{z}{r}\right)^2 
\right]
\psi^{ -\frac{3D-5}{D-3} } = 0 \ .
\end{aligned}
\end{equation}
It turns out to be convenient for solving this differential
equation to
introduce a hyperspherical coordinate system on the $D-1$-dimensional spatial slices, such that the flat conformal metric is
\begin{align*}
d\hat s^2 &= \hat{\gamma}_{ab} dx^a dx^b \\
          &= dr^{2} + r^2 \left[ d\vartheta^2 + \sin \vartheta^2
              \left(d\varphi^2
            +  \sin^2 \varphi d\Omega_{D-4}\right)\right] \ ,
\end{align*}
with $ \cos \vartheta = \frac{z}{r}$. We further introduce the radial
coordinate
%
\begin{equation}
  \label{eq:Adef}
  X \equiv \left(
  1 + \frac{\mu}{4 r^{D-3}}
  \right)^{-1} \, ,
\end{equation}
which reduces to the coordinate $A$ of Eq.~(31) in~\cite{Ansorg:2004ds}
for the case of $D=4$ spacetime dimensions.
Expressed in the new coordinate system, Eq.~\eqref{eq:ueqsingle} becomes
\begin{widetext}
  \begin{equation}
    \label{eq:solveu}
    \begin{aligned}
    &  \left\{
        \partial_{XX} + \frac{2}{X} \partial_X + \frac{1}{(D-3)^2 X^2 (1-X)^2} \left[
          \partial_{\vartheta \vartheta} + (D-3) \cot \vartheta \partial_{\vartheta}
          + \frac{1}{\sin^2 \vartheta} \left(
            \partial_{\varphi \varphi} + (D-4) \cot \varphi \partial_{\varphi}
          \right)
        \right]
      \right\} u  \\
     & = -\alpha \left( \frac{P_z}{\mu} \right)^2 X^{-\frac{D-7}{D-3}} 
     \left(
       1 + u X
     \right)^{-\frac{3D-5}{D-3}}
     \left(1 + \frac{D(D-3)}{2} \cos^2 \vartheta \right) \ ,
\end{aligned}
\end{equation}
\end{widetext}
with 
\[\alpha \equiv \frac{128 \pi^2 (D-1)^2}{(D-3)(D-2)^3 \mathcal{A}^2_{D-2}} \ .\]
For $D=4$, we recover Eq.~(40) of~\cite{Ansorg:2004ds}.
In order to study the behavior of the solution at spatial infinity, we now
perform a Taylor expansion in $v \equiv \frac{P_z}{\mu}$,
\begin{equation}
  \label{eq:7}
  u = \sum_{j=1}^{\infty} v^{2j} u_j \,.
\end{equation}
Odd powers of $v$ have to vanish in order to satisfy Eq.~(\ref{eq:solveu}). We have the following equation for $u_1$
\begin{widetext}
\begin{equation}
  \label{eq:solveu1}
  \begin{aligned}
    &  \left\{
       \partial_{XX} + \frac{2}{X} \partial_X + \frac{1}{(D-3)^2 X^2 (1-X)^2}
       \left[
        \partial_{\vartheta \vartheta} + (D-3) \cot \vartheta
             \partial_{\vartheta}
      \right]
    \right\} u_1 =
     -\alpha  X^{-\frac{D-7}{D-3}}
    \left(1 + {\textstyle \frac{D(D-3)}{2} } \cos^2 \vartheta \right) \, .
  \end{aligned}
\end{equation}
\end{widetext}
In order to solve Eq.~\eqref{eq:solveu1}, we make the \emph{ansatz}
\begin{equation}
\label{eq:12}
 u_1 = f(X) + g(X) Q_D(\cos \vartheta) \ ,
\end{equation}
where $Q_D (\cos \vartheta ) = (D-1) \cos^2 \vartheta - 1$. By solving Eq.~(\ref{eq:solveu1}), we find that
the functions $f(X)$ and $g(X)$ take the form
\begin{align}
\label{eq:fg}
f(X)  = \frac{32 \pi^2 (D-3) }{(D-2)^2 \mathcal{A}^2_{D-2}}
\left(  1 - X^{\frac{D+1}{D-3}}\right) \ ,
\end{align}
%
\begin{align}
 & g(X)  = \nonumber \\
& k_1 \left( \frac{X}{1-X} \right)^{\frac{2}{D-3}} +
  k_{2} \left( \frac{1-X}{X} \right)^{\frac{D-1}{D-3}} 
   - \alpha {\textstyle \frac{D(D-3)^3}{2(D+1)(D-1)} }\times \nonumber \\
 & \times\Bigg[ {\textstyle \frac{1}{D-1}}
  \frac{X^{\frac{D+1}{D-3}}}{(1-X)^{\frac{2}{D-3}}} {}_2 F_1 \left(
    {\textstyle -\frac{D-1}{D-3}, \frac{D-1}{D-3}; 2 \frac{D-2}{D-3};} X
  \right) \nonumber  \\
&   - {\textstyle \frac{1}{2D} } X^{\frac{D+1}{D-3}} (1-X)^{\frac{D-1}{D-3}} 
  {}_2 F_1 \left( 
    {\textstyle  \frac{2}{D-3}, \frac{2D}{D-3}; 3 \frac{D-1}{D-3};} X \right) \Bigg] \ ,
\end{align}
%
where ${}_2 F_1(a,b;c;X)$ is the hypergeometric function and
$k_{1,2}$ 
are constants to be fixed by imposing that $g(X=1) = 0$
and $g(X=0)$ is smooth.  Requiring analyticity at $X=0$ and using the
property $F(a,b,c,0)=1$, we immediately find $k_2=0$.

We are now interested in the large $X\to 1$ limit. 
Therefore, we use the $z \rightarrow 1-z$ transformation law for the hypergeometric functions~\cite{abramowitz},
\begin{widetext}
\begin{align}
F(a\!-\!c\!+\!1,b\!-\!c\!+\!1,2\!-\!c,z) & =
(1\!-\!z)^{c-a-b}
\frac{\Gamma(2-c)\Gamma(a+b-c)}{\Gamma(a-c+1)\Gamma(b-c+1)}
 \,F(1\!-\!a,1\!-\!b,c\!-\!a\!-\!b\!+\!1,1\!-\!z)  \nonumber \\
& \quad + \frac{\Gamma(2-c)\Gamma(c-a-b)}{\Gamma(1-a)\Gamma(1-b)}
 \,F(a\!-\!c\!+\!1,b\!-\!c\!+\!1,-c\!+\!a\!+\!b\!+\!1,1\!-\!z)\ .
\end{align}
\end{widetext}
Requiring a regular solution, we find that $k_1$ has to satisfy
\begin{equation}
k_1=\frac{64 \pi^2 D(D-3)^2}{(D-2)^3(D+1)  \mathcal{A}^2_{D-2}}
\frac{\Gamma\left(\frac{2(D-2)}{D-3}\right)^2}{\Gamma\left(\frac{3D-5}{D-3}\right)}\label{eq:13} \ .
\end{equation}

Let us write these functions explicitly for 
$D=4,5,7$ (for $D=6$ the hypergeometric function does not simplify):
\begin{itemize}
\item $D=4:$
  \begin{align}
    \label{eq:fgD4}
    f(X) & = \frac{1}{2} \left( 1 - X^5 \right) \ , \\
    g(X) & = \frac{(1-X)^2}{10 X^3} \Big[
     84 (1-X) \log(1-X) \notag \\
  & \quad + 84X - 42 X^2 - 14 X^3 - 7 X^4 \notag \\
  & \quad -4 X^5 - 2 X^6 
  \Big] \ ;
  \end{align}
These are Eqs.~(42--44) in~\cite{Ansorg:2004ds}, with appropriate redefinitions.
\item $D=5:$
  \begin{align}
    \label{eq:fgD5}
    f(X) & = \frac {16}{9 \pi ^2} \left(1-X^3\right) \ , \\
    g(X) & = -\frac{80 (1-X)^2 }{81 \pi^2 X^2} 
    \Big[
      4 \log(1-X) \notag \\
      & \quad + 4X + 2 X^2+X^3
       \Big] \ ;
  \end{align}
\item $D=7:$
  \begin{align}
    \label{eq:fgD7}
    f(X) & = \frac{128 \left(1-X^2\right)}{25 \pi ^4} \ , \\
    g(X) & = \frac{28}{125 \pi^4 \sqrt{(1-X) X^3}}
    \Big[-30 \sqrt{(1-X) X} \notag \\
      & \quad  + 40 \sqrt{(1-X) X^3}-16 \sqrt{(1-X)  X^7} \notag \\
      & \quad +3 \pi X^2 +6 \left(5-10 X+4 X^2\right) \arcsin \sqrt{X} \Big] \ .
  \end{align}
\end{itemize}
Analyzing these expressions, we can anticipate the convergence properties of the numerical solutions obtained in terms of pseudo-spectral methods. 
For instance, analyticity of $f$ and $g$ suggests exponential convergence.
As will become clear in the next section, we are interested in the convergence properties in a coordinate $A$ behaving as $A \sim 1-\frac{1}{r}$, for large $r$. 
We thus introduce a coordinate $A$ that satisfies
\begin{equation}
\label{eq:XtoA}
X=\left(
  1 + (A^{-1}-1)^{D-3}
\right)^{-1} \,.
\end{equation}
In terms of the $A$ coordinate, we find that the functions $f$ are analytical. 
For the function $g$ in the vicinity of $A = 1$, the leading terms behave as follows:
\begin{itemize}
\item $D=5$
  \begin{align}
    \label{eq:singD5}
   g(A) \sim -\frac{80}{81\pi^2} (1-A)^4 \left[ 8\log(1-A) + 7 \right] \,, 
  \end{align}
\item $D=6$
  \begin{align}
    \label{eq:singD6}
   g(A) \sim \frac{19683}{6272\pi^2} (1-A)^5 \,,
  \end{align}
\item $D=7$
  \begin{align}
    \label{eq:singD7}
   g(A) \sim \frac{84}{25\pi^3} (1-A)^6 \,.
  \end{align}
\end{itemize}
From the behaviour of the functions $f$ and $g$ and Eq.~(\ref{eq:12})
we conclude that the first term in the expansion
(\ref{eq:7}) has a leading-order behaviour $u_1 \sim 1/r^{D-3}$
as $r\rightarrow \infty$. 
Iteratively solving Eq.~\eqref{eq:solveu} for higher powers of $v$ is complicated by the presence of the source terms on the right-hand side, but under simplifying assumptions indicates that higher-order terms $u_j \ge 2$ acquire additional factors of $1/r$ and therefore the leading-order falloff behaviour is given correctly by that of $u_1$. This result is confirmed by our numerical investigation using finite boost parameters as we shall discuss in the next section.

With regard to the analyticity of the solutions and the resulting expectations
for the convergence properties of a spectral algorithm, we summarize the results
of our analytical study of a single puncture as follows.  In $D=6,7$, the
leading terms are analytic functions in the vicinity of $A=1$.  Actually, for
$D=7$, $g(A)$ is analytic in the vicinity of any point. Therefore, we expect
exponential convergence of the pseudo-spectral code.  For $D=5$, one observes
the presence of a logarithmic term.  This type of term is known to arise in
$D=4$, when punctures have nonvanishing
momenta~\cite{Dain:2001ry,Gleiser:1999hw} and in that case their presence makes
the convergence algebraic in the single puncture case. In the next section we
shall investigate the impact of the logarithmic terms on the convergence
properties of our spectral solver.

\section{Two punctures with linear momentum}
\label{sec5}
\subsection{Code changes}
We first explicitly list the modifications applied to the spectral
solver of Ref.~\cite{Ansorg:2004ds} and demonstrate how these modifications
enable us to generate initial data for boosted black hole binaries with
convergence properties and levels of constraint violation similar to the
$D=4$ case.
For this purpose we start by recalling that the spectral
solver of \cite{Ansorg:2004ds} employs coordinates
\begin{equation}
\label{eq:ABphi}
A \in [0,1]\ , \quad B \in [-1,1]\ , \quad \phi \in [0,2\pi] \ ,
\end{equation}
which are defined by Eq.~(62) of~\cite{Ansorg:2004ds},
\begin{equation}
\begin{split}
x&=b\frac{2A}{1-A^2}\frac{1-B^2}{1+B^2}\sin\phi\,, \\
y&=b\frac{2A}{1-A^2}\frac{1-B^2}{1+B^2}\cos\phi\,,\\
z&=b\frac{A^2+1}{A^2-1}\frac{2B}{1+B^2}\,,
\end{split}
\end{equation}
where $b$ is half of the coordinate distance between the punctures.
In particular, the coordinate $A$ satisfies
\begin{equation}
  r\to\infty \Leftrightarrow A\to 1\label{eq:rinfty} \ .
\end{equation}

The first modification consists in adapting the source term and Laplace operator
according to~\eqref{eq:elleq0}.

Next, we note that the type of high-energy collisions which form the main
motivation for this work often start from relatively large initial separations
of the holes, $|z_1-z_2|\gg r_S$.
In order to obtain high-precision solutions for such binary configurations,
we found it crucial to introduce a coordinate $A'$ defined as
\begin{equation}
\label{eq:barA}
A = \frac{\sinh\left[\kappa (A'+ 1 )/2 \right]}{\sinh{\kappa}} \ ,
\end{equation}
where $\kappa$ is an adjustable free parameter.
Note that for $\kappa = 0$ we obtain $A = \frac{1}{2}(A' + 1)$
For $\kappa > 0$, however, the new coordinate $A'$ provides the spectral method
with enhanced resolution near $A \sim 0$.

A further modification is related to the asymptotic falloff of the
function $u$ as obtained in the previous section,
\begin{equation}
  \label{eq:ufalloff}
  u \sim \frac{1}{r^{D-3}} \  .
\end{equation}
To naturally accommodate this behaviour with the spectral
coordinates used in the code, we have changed the variable $U$ of Eq.~(5)
in~\cite{Ansorg:2004ds} to
\begin{equation}
  \label{eq:U}
  u = (A' - 1)^{D-3} U \ .
\end{equation}
Note that this $U$ variable is the variable that the code actually solves for.


Finally, we adjust the calculation of the ADM mass from the numerical
solution. For this purpose, we note that, asymptotically
\begin{equation}
  \label{eq:psiasympt}
  \psi = 1 + \frac{\mu_{+}}{4r_{+}^{D-3}} + \frac{\mu_{-}}{4r_{-}^{D-3}} + u 
  \sim 1 + \frac{\mu}{4r^{D-3}} \ ,
\end{equation}
with $\mu \equiv r_{S_{\rm global}}^{D-3} \equiv \frac{16\pi
M_{\mathrm{ADM}}}{\mathcal{A}_{D-2}(D-2)}$ and $\mu_{\pm} \equiv r_{S_{(\pm)}}^{D-3}$. The ADM mass is then obtained
from
\begin{align}
  \label{eq:rsglobal}
  r_{S_{\rm global}}^{{D-3}} & = r_{S_{(+)}}^{D-3} + r_{S_{(-)}}^{D-3}
      + 4 \lim_{r\to\infty} r^{D-3} u \notag \\
            & = r_{S_{(+)}}^{D-3} + r_{S_{(-)}}^{D-3} + 4
            \left(-2b \frac{\tanh \kappa}{\kappa} \right)^{D-3}
            U(A' = 1) \ ,
\end{align}
where we have used Eq.~(62) of~\cite{Ansorg:2004ds}, and Eq.~\eqref{eq:barA}
and~\eqref{eq:U}. We show in Table~\ref{tab:admmass} the values obtained for the ADM mass of some cases we considered.
\begin{table}
\begin{tabular}{c c c c c}
  \hline
  $D$ & $b/r_S$ & $P/r_{S}^{D-3}$ & $r_{S_{\rm global}}^{{D-3}}/r_{S}^{D-3}$ & $M_{\rm ADM}/r_{S}^{D-3}$ \\
  \hline  \hline
  4   & 30.185   & 0.8          &  3.555          & 1.78       \\
  \hline
  5   & 30.185   & 0.8          &  1.931          & 2.27       \\
  \hline
  6  &  30.185   & 0.8          &  1.415          & 2.96       \\
  \hline
  7  & 30.185    & 0.8          &  1.236          & 3.81       \\
  \hline
\end{tabular}
\caption{\label{tab:admmass} ADM mass obtained with Eq.~\eqref{eq:rsglobal} in units of the ``bare'' Schwarzschild radius $r_S^{D-3} = r_{S_{(+)}}^{D-3} + r_{S_{(-)}}^{D-3}$. 
The variation of the ADM mass with resolution is of the order of $10^{-10}$ for all $D$ and $n \ge 100$ grid points indicating that the accuracy in the ADM mass is limited by round-off errors.}
\end{table}

\subsection{Results}
We now study the numerical results as obtained for $D=4,5,6,7$
with these adaptations
of the spectral solver of \cite{Ansorg:2004ds}. Throughout the remainder
of this section we will graphically present results in units of the
``bare'' Schwarzschild radius defined as
$r_S^{D-3} = r_{S_{(+)}}^{D-3} + r_{S_{(-)}}^{D-3}$.

\begin{figure}[htbp]
\centerline{\includegraphics[clip=true,width=0.5\textwidth]{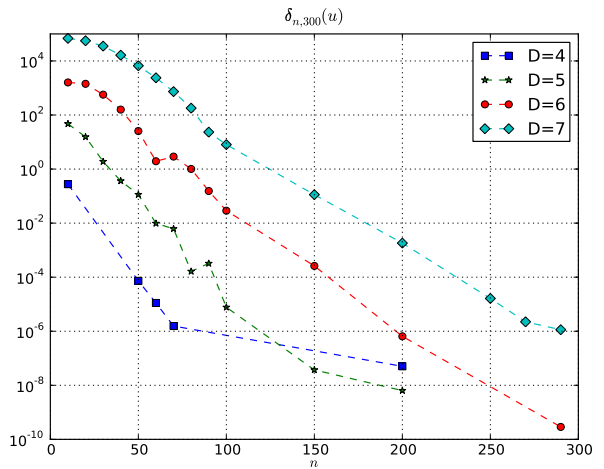}}
\caption[]{\label{fig:d60_conv} 
Convergence plot for the $b/r_S=30.185$, $P/r_{S}^{D-3}=\pm0.80$ cases.
}
\label{convergenceplot}
\end{figure}

We first address the convergence properties of the numerical algorithm
by evaluating the quantity
\begin{equation}
  \label{eq:conv}
  \delta_{n,m}(u) = \max |1-u_n/u_m| \,,
\end{equation}
where the maximum is obtained along the collision axis, \emph{i.e.}~$z$-axis
in our case. Here, the index $m$ refers to a reference solution obtained using
a large number $m$ of grid points while $n$ denotes test solutions using
a coarser resolution, $n<m$. The result obtained for black hole binaries
with initial separation $b/r_S=30.185$ and boost $P^z/r_S^{D-3}=0.8$
in $D=4$, $5$, $6$ and $7$ dimensions
is displayed in Fig. \ref{convergenceplot}. We note from this
figure, that achieving a given target accuracy $\delta_{n,m}$
requires a larger number of points $n$ as $D$ increases. We
emphasize in this context, however, that this increase in computational
cost in higher dimensions is unlikely to significantly affect the total
computational cost of the simulations which typically are dominated by
the time evolution rather than the initial data calculation.
Most importantly, we observe exponential convergence up to a level
of $\delta_{n,m}(u) \approx 10^{-6}$ for all values of the
spacetime dimensionality $D$. Below that level, the two leftmost curves
in Fig.~\ref{fig:d60_conv}, 
corresponding to $D=4$ and $D=5$, respectively,
show that the rate of convergence decreases indicating
that the logarithmic terms become significant and reduce the convergence
to algebraic level similar to the observation in Fig.~4 of
Ref.~\cite{Ansorg:2004ds}. For $D=6$, the convergence remains exponential,
in agreement with the absence of logarithmic terms in the analysis
of Sec.~\ref{sec4}. Irrespective of a change to algebraic convergence,
however, our algorithm is capable of reducing the quantity $\delta_{m,n}(u)$
for all values of $D$ to a level comparable to the case $D=4$ and, thus,
producing initial data of similar quality as in 3+1 dimensions, provided
we use a sufficiently high resolution $n$.

\begin{figure}[htbp]
\centerline{\includegraphics[clip=true,width=0.5\textwidth]{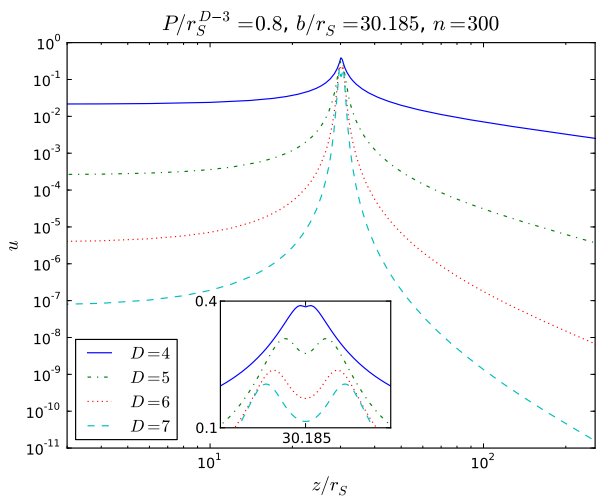}}
\caption[]{\label{fig:u_loglog} $u$ function for $D=4,\dots,7$ plotted along the $z$-axis, in units of $ r_S $. We used $n_A = n_B = n = 300$, $n_{\phi}=4$. We also show a zoom around the puncture.}
\end{figure}

For illustration, we plot
in Fig.~\ref{fig:u_loglog} the function $u$ obtained for the case of
$b/r_S=30.185$, $P^z/r_{S}^{D-3}=0.8$. The behaviour is qualitatively similar for all values of $D$,
but the figure demonstrates the faster falloff for larger $D$ as predicted by \eqref{eq:ufalloff}. For this plot we have used
$n_A=300$, $n_B=300$ and $n_\phi=4$ grid points.
The inset in the figure shows the function $u$ in the immediate
vicinity of the puncture. While the profile develops multiple
extrema for $D>4$, the profile remains smooth for all values of $D$.

\begin{figure}[htbp]
\centerline{\includegraphics[clip=true,width=0.5\textwidth]{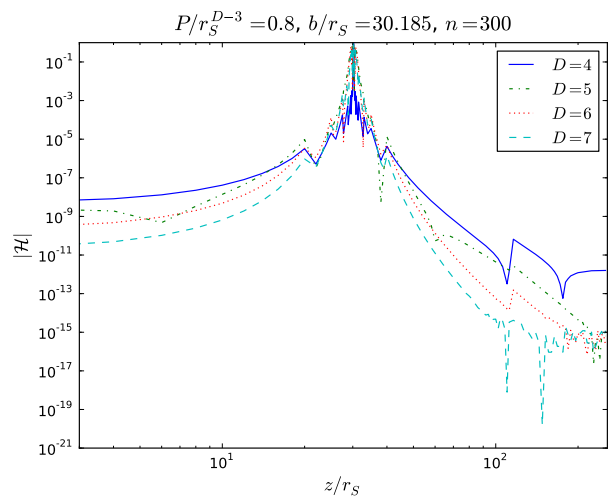}}
\caption[]{\label{fig:hc} Violation of the Hamiltonian constraint along
the $z$-axis, evaluated with a fourth-order finite-difference scheme. The
growth of the constraint violation near the puncture is an artifact of
finite-differencing across the puncture; see text for details.}
\end{figure}

Finally, we show in Fig.~\ref{fig:hc}, the Hamiltonian constraint
corresponding to the solutions presented in Fig.~\ref{fig:u_loglog}
as measured by a fourth-order
finite-differencing scheme of the evolution code \cite{Sperhake:2006cy}.
We emphasize that the violation of Eq.~(\ref{eq:u}) inside the spectral
initial data solver is $<10^{-12}$ by construction. The independent
evaluation of the constraint violation in the evolution code
serves two purposes. First, it checks
that the differential Eq.~(\ref{eq:u}) solved by the spectral method
corresponds to the Hamiltonian constraint formulated in ADM variables;
an error in coding up the differential Eq.~(\ref{eq:u}) could still
result in a solution for $u$ of the spectral solver, but would manifest
itself in significantly larger violations in Fig.~\ref{fig:hc}. Second,
it demonstrates that the remaining numerical error
is dominated by the time evolution instead of the initial solver.
Note in this context that the relatively large violations of order
unity near the
puncture location in Fig.~\ref{fig:hc} are an artifact of the
fourth-order discretization in the diagnostics of the evolution code
and are typical for evolutions of the moving-puncture type;
see e.~g.~the right panel in Fig.~(8) in Brown {\em et al.} \cite{Brown2009}.

The solid (blue) curve
obtained for the ``standard'' $D=4$ case serves as reference.
For all values of $D$, the constraint violations are maximal at
the puncture location $z_1/r_S\approx 15$ and rapidly decrease
away from the puncture. As expected from the higher falloff rate
of the grid functions for larger $D$, the constraints also
drop faster for higher-dimensionality of the spacetime.
%
\section{Conclusions}
In this paper we have presented numerical solutions of the Einstein constraint
equations for the construction of initial data containing single or binary black
holes with nonvanishing linear momentum in $D>4$ dimensional spacetimes. For
this purpose we have modified the spectral solver of
Ref.~\cite{Ansorg:2004ds}. As in $D=4$ dimensions, the momentum constraints
decouple from the Hamiltonian constraint under the assumption of conformal
flatness and a spatially constant trace of the extrinsic curvature and allow for
an analytic solution describing multiple black holes with nonzero momenta. One
thus arrives at a single elliptic differential equation for the conformal factor
or, to be more specific, a regular correction function $u$ to the
Brill-Lindquist part of the conformal factor. We have studied the resulting
differential equation in the limit of a single black hole with small boost in
order to investigate the asymptotic behaviour of $u$ at spatial infinity, where
we find $u\sim 1/r^{D-3}$.  For $D=6,7$ we further observe that $u$ is an analytic
function expanded in terms of the coordinate $A$ [cf.~Eq.~(\ref{eq:XtoA})]
around spatial infinity, so that a spectral algorithm should provide exponential
convergence. For $D=5$ the expansion of $u$ includes a logarithmic term, but, as
has also been observed in Ref.~\cite{Ansorg:2004ds} for $D=4$, this term
is subdominant in the case of a
black hole binary with equal and opposite momenta. We have used the asymptotic
behaviour of $u$ to adapt the set of coordinates employed in the spectral solver
to arbitrary dimensionality and further performed a transformation to a radial
coordinate that ensures sufficient resolution near spatial infinity for the case
of large separation of the two holes.

The resulting code has been used to calculate initial data for
black hole binaries with linear momenta along the $z$-axis,
\emph{i.e.}~corresponding to a head-on collision. Even though the number
of grid points required for reaching a given threshold accuracy increases
with $D$, we observe rapid convergence for all values $D=4,5,6,7$.
Although for large resolutions $n$ the convergence becomes algebraic
due to logarithmic terms in the $r$ dependency of the solution $u$
for $D=4$ and $D=5$, this transition can be compensated with
a moderate increase in the number of grid points $n$ as has also been
observed in Ref.~\cite{Ansorg:2004ds} for $D=4$.
Closer investigation of the profile of the function $u$ thus obtained
confirms the expected higher falloff rate as $r\rightarrow \infty$
for larger $D$. We further note that $u$ shows smooth behaviour near
the puncture. Finally, we have studied the Hamiltonian constraint
as a function along the collision axis. As in $D=4$, the residual
constraint violations after the elliptic solving are largest near
the puncture and rapidly falloff away from the puncture. As expected
from the asymptotic behaviour of $u$, the constraint violations
decay even faster away from the puncture as $D$ increases.

The construction of initial data forms a crucial step in performing
high-energy collisions of black hole binaries in higher-dimensional
spacetimes which will complement existing studies in $D=4$ dimensions
\cite{Sperhake:2008ga,Shibata:2008rq,Sperhake:2009jz} as well as studies
in $D=5$ dimensions \cite{Okawa:2011fv} starting from superposed
single black hole initial data.

\begin{acknowledgments}
  We thank Andrea Nerozzi and the anonymous referee for useful suggestions and
  discussions.  M.Z. and H.W. are funded by FCT through grants
  SFRH/BD/43558/2008 and SFRH/BD/46061/2008.  U.S. acknowledges support from the
  Ram\'on y Cajal Programme of the Ministry of Education and Science of Spain,
  the FP7-PEOPLE-2011-CIG Grant CBHEO, Number 293412, NSF grants PHY-0601459,
  PHY-0652995 and the Sherman Fairchild Foundation to Caltech.  This work was
  supported by the {\it DyBHo--256667} ERC Starting Grant and by FCT - Portugal
  through projects PTDC/FIS/098025/2008, PTDC/FIS/098032/2008,
  PTDC/CTE-AST/098034/2008, CERN/FP/116341/2010.  This research was supported by
  an allocation through the TeraGrid Advanced Support Program under grant
  PHY-090003, an allocation by the Centro de Supercomputaci{\'o}n de Galicia
  (CESGA) under project ICTS-2009-40 and allocations at the Barcelona
  Supercomputing Center (BSC) under projects AECT-2011-2-0006 and
  AECT-2011-2-0015.  Computations were performed on the TeraGrid clusters Kraken
  at the National Institute for Computational Sciences (NICS) of the University
  of Tennessee and Trestles at the San Diego Supercomputing Center (SDSC), the
  Milipeia cluster in Coimbra, Finis Terrae at the Supercomputing Center of
  Galicia (CESGA), the supercomputer Caesaraugusta at the Institute for
  Biocomputation and Physics of Complex Systems (BIFI) at the University of
  Zaragoza, HLRB2 of the Landesrechenzentrum (LRZ) in Munich, MareNostrum at the
  Barcelona Supercomputing Center in Barcelona and on the Blafis cluster at the
  University of Aveiro.
  
\end{acknowledgments}

\providecommand{\href}[2]{#2}

\begingroup\begin{thebibliography}{10}

\bibitem{Ansorg:2004ds}
M.~Ansorg, B.~Bruegmann, and W.~Tichy, ``{A single-domain spectral method for
  black hole puncture data},''
  \href{http://dx.doi.org/10.1103/PhysRevD.70.064011}{{\em Phys. Rev.} {\bf
  D70} (2004)  064011},
\href{http://arxiv.org/abs/gr-qc/0404056}{{\tt arXiv:gr-qc/0404056}}.

\bibitem{Antoniadis:1990ew}
I.~Antoniadis, ``{A Possible new dimension at a few TeV},''
\href{http://dx.doi.org/10.1016/0370-2693(90)90617-F}{{\em Phys. Lett.} {\bf
  B246} (1990)  377--384}.

\bibitem{Antoniadis:1998ig}
I.~Antoniadis, N.~Arkani-Hamed, S.~Dimopoulos, and G.~R. Dvali, ``{New
  dimensions at a millimeter to a Fermi and superstrings at a TeV},''
  \href{http://dx.doi.org/10.1016/S0370-2693(98)00860-0}{{\em Phys. Lett.} {\bf
  B436} (1998)  257--263},
\href{http://arxiv.org/abs/hep-ph/9804398}{{\tt arXiv:hep-ph/9804398}}.

\bibitem{Randall:1999ee}
L.~Randall and R.~Sundrum, ``{A large mass hierarchy from a small extra
  dimension},'' \href{http://dx.doi.org/10.1103/PhysRevLett.83.3370}{{\em Phys.
  Rev. Lett.} {\bf 83} (1999)  3370--3373},
\href{http://arxiv.org/abs/hep-ph/9905221}{{\tt arXiv:hep-ph/9905221}}.

\bibitem{Argyres:1998qn}
P.~C. Argyres, S.~Dimopoulos, and J.~March-Russell, ``{Black holes and
  sub-millimeter dimensions},''
  \href{http://dx.doi.org/10.1016/S0370-2693(98)01184-8}{{\em Phys. Lett.} {\bf
  B441} (1998)  96--104},
\href{http://arxiv.org/abs/hep-th/9808138}{{\tt arXiv:hep-th/9808138}}.

\bibitem{Banks:1999gd}
T.~Banks and W.~Fischler, ``{A model for high energy scattering in quantum
  gravity},''
\href{http://arxiv.org/abs/hep-th/9906038}{{\tt arXiv:hep-th/9906038}}.

\bibitem{Giddings:2001bu}
S.~B. Giddings and S.~D. Thomas, ``{High energy colliders as black hole
  factories: The end of short distance physics},''
  \href{http://dx.doi.org/10.1103/PhysRevD.65.056010}{{\em Phys. Rev.} {\bf
  D65} (2002)  056010},
\href{http://arxiv.org/abs/hep-ph/0106219}{{\tt arXiv:hep-ph/0106219}}.

\bibitem{Dimopoulos:2001hw}
S.~Dimopoulos and G.~L. Landsberg, ``{Black Holes at the LHC},''
  \href{http://dx.doi.org/10.1103/PhysRevLett.87.161602}{{\em Phys. Rev. Lett.}
  {\bf 87} (2001)  161602},
\href{http://arxiv.org/abs/hep-ph/0106295}{{\tt arXiv:hep-ph/0106295}}.

\bibitem{Ahn:2002mj}
E.-J. Ahn, M.~Cavaglia, and A.~V. Olinto, ``{Brane factories},''
  \href{http://dx.doi.org/10.1016/S0370-2693(02)03011-3}{{\em Phys. Lett.} {\bf
  B551} (2003)  1--6},
\href{http://arxiv.org/abs/hep-th/0201042}{{\tt arXiv:hep-th/0201042}}.

\bibitem{Chamblin:2004zg}
A.~Chamblin, F.~Cooper, and G.~C. Nayak, ``{SUSY production from TeV scale
  blackhole at LHC},'' \href{http://dx.doi.org/10.1103/PhysRevD.70.075018}{{\em
  Phys. Rev.} {\bf D70} (2004)  075018},
\href{http://arxiv.org/abs/hep-ph/0405054}{{\tt arXiv:hep-ph/0405054}}.

\bibitem{Kanti:2008eq}
P.~Kanti, ``{Black Holes at the LHC},''
  \href{http://dx.doi.org/10.1007/978-3-540-88460-6_10}{{\em Lect. Notes Phys.}
  {\bf 769} (2009)  387--423},
\href{http://arxiv.org/abs/0802.2218}{{\tt arXiv:0802.2218 [hep-th]}}.

\bibitem{Yoshino:2009xp}
H.~Yoshino and M.~Shibata, ``{Higher-dimensional numerical relativity:
  Formulation and code tests},''
  \href{http://dx.doi.org/10.1103/PhysRevD.80.084025}{{\em Phys. Rev.} {\bf
  D80} (2009)  084025},
\href{http://arxiv.org/abs/0907.2760}{{\tt arXiv:0907.2760 [gr-qc]}}.

\bibitem{Sorkin:2009bc}
E.~Sorkin and M.~W. Choptuik, ``{Generalized harmonic formulation in spherical
  symmetry},''
\href{http://arxiv.org/abs/0908.2500}{{\tt arXiv:0908.2500 [gr-qc]}}.

\bibitem{Zilhao:2010sr}
M.~Zilhao {\em et al.}, ``{Numerical relativity for D dimensional axially
  symmetric space-times: formalism and code tests},''
  \href{http://dx.doi.org/10.1103/PhysRevD.81.084052}{{\em Phys. Rev.} {\bf
  D81} (2010)  084052},
\href{http://arxiv.org/abs/1001.2302}{{\tt arXiv:1001.2302 [gr-qc]}}.

\bibitem{Lehner2010}
L.~Lehner and F.~Pretorius, ``{Black Strings, Low Viscosity Fluids, and
  Violation of Cosmic Censorship},'' {\em Phys. Rev. Lett.} {\bf 105} (2010)
  101102. arXiv:1006.5960 [hep-th].

\bibitem{Witek:2010xi}
H.~Witek, M.~Zilhao, L.~Gualtieri, V.~Cardoso, C.~Herdeiro, {\em et al.},
  ``{Numerical relativity for D dimensional space-times: head-on collisions of
  black holes and gravitational wave extraction},''
  \href{http://dx.doi.org/10.1103/PhysRevD.82.104014}{{\em Phys.Rev.} {\bf D82}
  (2010)  104014}, \href{http://arxiv.org/abs/1006.3081}{{\tt arXiv:1006.3081
  [gr-qc]}}.

\bibitem{Witek:2010az}
H.~Witek, V.~Cardoso, L.~Gualtieri, C.~Herdeiro, U.~Sperhake, {\em et al.},
  ``{Head-on collisions of unequal mass black holes in D=5 dimensions},''
  \href{http://dx.doi.org/10.1103/PhysRevD.83.044017}{{\em Phys.Rev.} {\bf D83}
  (2011)  044017}, \href{http://arxiv.org/abs/1011.0742}{{\tt arXiv:1011.0742
  [gr-qc]}}.

\bibitem{Okawa:2011fv}
H.~Okawa, K.-i. Nakao, and M.~Shibata, ``{Is super-Planckian physics visible?
  -- Scattering of black holes in 5 dimensions},''
  \href{http://dx.doi.org/10.1103/PhysRevD.83.121501}{{\em Phys.Rev.} {\bf D83}
  (2011)  121501}, \href{http://arxiv.org/abs/1105.3331}{{\tt arXiv:1105.3331
  [gr-qc]}}.

\bibitem{Sperhake:2008ga}
U.~Sperhake, V.~Cardoso, F.~Pretorius, E.~Berti, and J.~A. Gonzalez, ``{The
  high-energy collision of two black holes},''
  \href{http://dx.doi.org/10.1103/PhysRevLett.101.161101}{{\em Phys. Rev.
  Lett.} {\bf 101} (2008)  161101},
\href{http://arxiv.org/abs/0806.1738}{{\tt arXiv:0806.1738 [gr-qc]}}.

\bibitem{Shibata:2008rq}
M.~Shibata, H.~Okawa, and T.~Yamamoto, ``{High-velocity collision of two black
  holes},'' \href{http://dx.doi.org/10.1103/PhysRevD.78.101501}{{\em Phys.
  Rev.} {\bf D78} (2008)  101501},
\href{http://arxiv.org/abs/0810.4735}{{\tt arXiv:0810.4735 [gr-qc]}}.

\bibitem{Sperhake:2009jz}
U.~Sperhake {\em et al.}, ``{Cross section, final spin and zoom-whirl behavior
  in high- energy black hole collisions},''
  \href{http://dx.doi.org/10.1103/PhysRevLett.103.131102}{{\em Phys. Rev.
  Lett.} {\bf 103} (2009)  131102},
\href{http://arxiv.org/abs/0907.1252}{{\tt arXiv:0907.1252 [gr-qc]}}.

\bibitem{Sperhake2010}
U.~Sperhake, E.~Berti, V.~Cardoso, F.~Pretorius, and N.~Yunes, ``{Superkicks in
  Ultrarelativistic Grazing Collisions of Spinning Black Holes},'' {\em Phys.
  Rev. D} {\bf 83} (2011)  024037. arXiv:1011.3281 [gr-qc].

\bibitem{Arnowitt:1962hi}
R.~Arnowitt, S.~Deser, and C.~W. Misner, ``{The dynamics of general
  relativity},''
\href{http://arxiv.org/abs/gr-qc/0405109}{{\tt arXiv:gr-qc/0405109}}.

\bibitem{York1979}
J.~W. {York}, Jr., ``{Kinematics and dynamics of general relativity},'' in {\em
  Sources of Gravitational Radiation}, {L.~L.~Smarr}, ed., pp.~83--126.
\newblock 1979.

\bibitem{Lichnerowicz1944}
A.~Lichnerowicz, ``L'integration des {\'e}quations de la gravitation
  relativiste et le probl{\`e}me des $n$ corps,'' {\em J. Math. Pures et Appl.}
  {\bf 23} (1944)  37--63.

\bibitem{York1971}
J.~W. York, Jr., ``Gravitational degrees of freedom and the initial-value
  problem,'' {\em Phys. Rev. Lett.} {\bf 26} (1971)  1656--1658.

\bibitem{York1972}
J.~W. York, Jr., ``Role of conformal three-geometry in the dynamics of
  gravitation,'' {\em Phys. Rev. Lett.} {\bf 28} (1972)  1082--1085.

\bibitem{York1973}
J.~W. York, Jr., ``Conformally invariant orthogonal decomposition of symmetric
  tensors on riemannian manifolds and the initial-value problem of general
  relativity,'' {\em J. Math. Phys.} {\bf 14} (1973)  456--464.

\bibitem{Cook:2000vr}
G.~B. Cook, ``{Initial Data for Numerical Relativity},'' {\em Living Rev. Rel.}
  {\bf 3} (2000)  5,
\href{http://arxiv.org/abs/gr-qc/0007085}{{\tt arXiv:gr-qc/0007085}}.

\bibitem{Alcubierre:2008}
M.~Alcubierre, {\em Introduction to 3+1 numerical relativity}.
\newblock International series of monographs on physics. Oxford Univ. Press,
  Oxford, 2008.

\bibitem{Bowen:1980yu}
J.~M. Bowen and J.~W. York~Jr., ``{Time asymmetric initial data for black holes
  and black hole collisions},''
\href{http://dx.doi.org/10.1103/PhysRevD.21.2047}{{\em Phys. Rev.} {\bf D21}
  (1980)  2047--2056}.

\bibitem{Brandt1997}
S.~Brandt and B.~Br{\"u}gmann, ``A simple construction of initial data for
  multiple black holes,'' {\em Phys. Rev. Lett.} {\bf 78} (1997)  3606--3609.

\bibitem{Campanelli2006}
M.~Campanelli, C.~O. Lousto, P.~Marronetti, and Y.~Zlochower, ``Accurate
  {E}volutions of {O}rbiting {B}lack-{H}ole {B}inaries without {E}xcision,''
  {\em Phys. Rev. Lett.} {\bf 96} (2006)  111101. gr-qc/0511048.

\bibitem{Baker2006}
J.~G. Baker, J.~Centrella, D.-I. Choi, M.~Koppitz, and J.~van Meter,
  ``Gravitational-{W}ave {E}xtraction from an inspiraling {C}onfiguration of
  {M}erging {B}lack {H}oles,'' {\em Phys. Rev. Lett.} {\bf 96} (2006)  111102.
  gr-qc/0511103.

\bibitem{Pretorius:2005gq}
F.~Pretorius, ``{Evolution of Binary Black Hole Spacetimes},''
  \href{http://dx.doi.org/10.1103/PhysRevLett.95.121101}{{\em Phys. Rev. Lett.}
  {\bf 95} (2005)  121101},
\href{http://arxiv.org/abs/gr-qc/0507014}{{\tt arXiv:gr-qc/0507014}}.

\bibitem{Scheel2008}
M.~A. Scheel, M.~Boyle, T.~Chu, L.~E. Kidder, K.~D. Matthews, and H.~P.
  Pfeiffer, ``{High-accuracy waveforms for binary black hole inspiral, merger,
  and ringdown},'' {\em Phys. Rev. D} {\bf 79} (2009)  024003. arXiv:0810.1767
  [gr-qc].

\bibitem{Yoshino:2006kc}
H.~Yoshino, T.~Shiromizu, and M.~Shibata, ``{Close-slow analysis for head-on
  collision of two black holes in higher dimensions: Bowen-York initial
  data},'' \href{http://dx.doi.org/10.1103/PhysRevD.74.124022}{{\em Phys. Rev.}
  {\bf D74} (2006)  124022},
\href{http://arxiv.org/abs/gr-qc/0610110}{{\tt arXiv:gr-qc/0610110}}.

\bibitem{Allen1999}
G.~Allen, T.~Goodale, J.~Mass\'o, and E.~Seidel, ``The cactus computational
  toolkit and using distributed computing to collide neutron stars,'' in {\em
  Proceedings of Eighth IEEE International Symposium on High Performance
  Distributed Computing, HPDC-8, Redondo Beach, 1999}.
\newblock IEEE Press, 1999.

\bibitem{cactus}
``{Cactus} {Computational} {Toolkit}.'' \url{http://www.cactuscode.org/}.

\bibitem{Shibata:1995we}
M.~Shibata and T.~Nakamura, ``{Evolution of three-dimensional gravitational
  waves: Harmonic slicing case},''
\href{http://dx.doi.org/10.1103/PhysRevD.52.5428}{{\em Phys. Rev.} {\bf D52}
  (1995)  5428--5444}.

\bibitem{Baumgarte:1998te}
T.~W. Baumgarte and S.~L. Shapiro, ``{On the numerical integration of
  Einstein's field equations},''
  \href{http://dx.doi.org/10.1103/PhysRevD.59.024007}{{\em Phys. Rev.} {\bf
  D59} (1999)  024007},
\href{http://arxiv.org/abs/gr-qc/9810065}{{\tt arXiv:gr-qc/9810065}}.

\bibitem{Brill:1963yv}
D.~R. Brill and R.~W. Lindquist, ``{Interaction energy in geometrostatics},''
\href{http://dx.doi.org/10.1103/PhysRev.131.471}{{\em Phys. Rev.} {\bf 131}
  (1963)  471--476}.

\bibitem{Yoshino:2005ps}
H.~Yoshino, T.~Shiromizu, and M.~Shibata, ``{The close limit analysis for
  head-on collision of two black holes in higher dimensions: Brill-Lindquist
  initial data},'' \href{http://dx.doi.org/10.1103/PhysRevD.72.084020}{{\em
  Phys. Rev.} {\bf D72} (2005)  084020},
\href{http://arxiv.org/abs/gr-qc/0508063}{{\tt arXiv:gr-qc/0508063}}.

\bibitem{abramowitz}
M.~Abramowitz and I.~A. Stegun, {\em Handbook of Mathematical Functions with
  Formulas Graphs and Mathematical Tables}.
\newblock Dover, New York, fifth~ed., 1964.

\bibitem{Dain:2001ry}
S.~Dain and H.~Friedrich, ``{Asymptotically flat initial data with prescribed
  regularity at infinity},''
  \href{http://dx.doi.org/10.1007/s002200100524}{{\em Commun.Math.Phys.} {\bf
  222} (2001)  569--609}, \href{http://arxiv.org/abs/gr-qc/0102047}{{\tt
  arXiv:gr-qc/0102047 [gr-qc]}}.

\bibitem{Gleiser:1999hw}
R.~J. Gleiser, G.~Khanna, and J.~Pullin, ``{Perturbative evolution of
  conformally flat initial data for a single boosted black hole},''
  \href{http://dx.doi.org/10.1103/PhysRevD.66.024035}{{\em Phys.Rev.} {\bf D66}
  (2002)  024035}, \href{http://arxiv.org/abs/gr-qc/9905067}{{\tt
  arXiv:gr-qc/9905067 [gr-qc]}}.

\bibitem{Sperhake:2006cy}
U.~Sperhake, ``{Binary black-hole evolutions of excision and puncture data},''
  \href{http://dx.doi.org/10.1103/PhysRevD.76.104015}{{\em Phys. Rev.} {\bf
  D76} (2007)  104015},
\href{http://arxiv.org/abs/gr-qc/0606079}{{\tt arXiv:gr-qc/0606079}}.

\bibitem{Brown2009}
D.~Brown, P.~Diener, O.~Sarbach, E.~Schnetter, and M.~Tiglio, ``{Turduckening
  black holes: An analytical and computational study},'' {\em Phys. Rev. D}
  {\bf 79} (2009)  044023. arXiv:0809.3533 [gr-qc].

\end{thebibliography}\endgroup
\end{document}